\documentstyle[12pt]{article}
\def\[{\left\lbrack}
\def\]{\right\rbrack}
 
\def\({\left(}
\def\){\right)}
\def\ih{\'\i}

\newcommand{\be}{\begin{equation}}
\newcommand{\ee}{\end{equation}}
\newcommand{\ea}{\end{eqnarray}}
\newcommand{\ba}{\begin{eqnarray}}

\title{Operatorial quantization of Born-Infeld Skyrmion model and hidden symmetries}

\author{J. Ananias Neto, C. Neves, E.R. de Oliveira \\and W. Oliveira
\thanks{e-mails:jorge,emanuel,cneves, wilson@fisica.ufjf.br.}\\
Departamento de F\ih sica, ICE \\ Universidade Federal de Juiz de
Fora, 36036-330, \\ Juiz de Fora, MG, Brazil\thanks{This work is 
supported in part by FAPEMIG, Brazilian Research Council.} }

\date{}

\begin{document}

\maketitle

\begin{abstract}
The SU(2) collective coordinates expansion of the Born-Infeld\break 
Skyrmion Lagrangian is performed. The classical
Hamiltonian is computed from this special Lagrangian 
in approximative way: it is derived from the expansion of this 
non-polynomial Lagrangian
up to second-order variable in the collective coordinates. 
This second-class constrained model is quantized by Dirac Hamiltonian 
method and symplectic formalism. Although it is not expected to find
symmetries on second-class systems, a hidden symmetry is disclosed by formulating the Born-Infeld Skyrmion model as a gauge theory. To this end we developed a new constraint conversion technique based on the symplectic formalism. Finally, a discussion on the role played by the hidden symmetry on the computation of the energy spectrum is presented.
\end{abstract}

\maketitle

\setlength{\baselineskip} {20 pt}

\vskip 1.5 cm

\section{Introduction}

The presence of symmetries in a constrained dynamical model reveals important physical contents of a given system. In particular, we can cite the energy spectrum that is an observable quantity invariant under gauge transformations. In views of this, some authors\cite{many} have proposed some processes to convert second-class systems to first-class ones.
In this paper, we are interested in investigating this subject employing a new technique based on the symplectic procedure, called symplectic gauge-invariant method\cite{ANO}. To clarify our proposal we will quantize a Skyrme-like model, discussed in this work.

The Skyrme model\cite{Skyrme} is an effective field 
theory for baryons and their interactions. These hadronic
particles are described from soliton solutions in the 
non-linear sigma model. Normally, in this Lagrangian, it is 
necessary to add the Skyrme term to stabilize the soliton solutions. 
In principle, the Skyrme term is arbitrary and there is not a concrete reason to fix it through a particular choice\cite{Weinberg}. 
Its importance
resides in the fact that, maybe, it is the most simple possible quartic 
derivative term that we need insert in the Hamiltonian in order  to
obtain soliton solutions\footnote{ According to Derrick scale 
theorem\cite{Derrick}.}. However, it is possible avoid this
ambiguity by adopting a nonconventional Lagrangian also based in a 
non-linear sigma model, given by

\begin{equation}
\label{BF}
L= - {F_\pi\over 16} \int d^3r \,  \[  Tr \partial_\mu U 
\partial^\mu U^+ \]^{3\over 2},
\end{equation}

\noindent where $F_\pi$ is the pion decay constant and $U$ is a SU(2) 
matrix. This model, based on the ideas of Born-Infeld 
Electrodynamics\cite{BInf}, was proposed by  Deser, Duff and
Isham\cite{Deser}. The existence of soliton solution might be 
observed \break by applying Derrick's theorem in the static 
Hamiltonian derived from the Lagrangian (\ref{BF}).

The main goal of this paper is to quantize the Born-Infeld Skyrmion model through the symplectic gauge-invariant method and then to display the role played by the hidden 
symmetry on the computation of the energy spectrum. To this end, this model is expanded
in terms of the collective rotational coordinates and, subsequently,
formulated as a gauge invariant theory. Implementing the semi-classical approach after the usual collective canonical
quantization, the spin and isospin modes are obtained producing quantum
corrections to the baryons properties\cite{Adkins}. This process reduces
the SU(2) Skyrme model to a nonrelativistic particle constrained
over a ${\cal{S}}^3$ sphere, a well known second-class problem\cite{sphere}\cite{NW3}.
Afterwards, the SU(2) Skyrme model expanded in terms of the collective rotational coordinates\cite{Adkins} is quantized via the Dirac Hamiltonian
method\cite{Dirac} and symplectic formalism\cite{FJ,JBW}, which allows
us to compute the field dependent Dirac's brackets among the physical
coordinates, assumed to be commutators at quantum level. We observe 
that when we keep the non-causality sector of the soliton solution influencing the physical values as minimum as possible, the commutators obtained are the same of the Skyrme model. At this level, problems 
involving operator ordering ambiguities\cite{HV,Toda} arise, which can be avoided just formulating this model as a gauge invariant theory. At this stage we unveil a hidden symmetry 
of the model, that is an unexpected result since this nonlinear model is originally a second-class model. 

For the sake of self consistency, this paper was organized as follows. In section 2, we propose the Born-Infeld 
Skyrmion model, 
obtaining the classical Hamiltonian by an approximative way from the expansion of the non-polynomial Lagrangian up to second-order variable 
in the collective coordinates since we take into account some causality arguments. In section 3, the second-class model will be quantized via 
Dirac and symplectic methods, where we demonstrate that the  
computed Dirac's brackets  are the same ones obtained for the Skyrme model\cite{Fujii,JAN}. In section 4, the Born-Infeld Skyrme model will be reformulated as a gauge theory 
via symplectic gauge-invariant method. In this section 
we will also investigate the hidden symmetry lying on the original phase-space coordinates. In order to corroborate the previous results, we investigate in section 5 the hidden 
symmetry via the gauge unfixing Hamiltonian method\cite{MR}. In section 6, the role played by the symmetry  on the computation of the energy spectrum will be explored. 
The last section is dedicated to the discussion of the physical meaning of our findings together with our final comments and conclusions. In appendices A and B,
we will present brief reviews about the new constraint conversion procedure, namely, the symplectic gauge-invariant formalism\cite{ANO}, and the gauge unfixing Hamiltonian formalism, respectively.

\section{The Born-Infeld Skyrmion model}

The dynamic system will be given performing the SU(2) collective
semi-classical expansion\cite{Adkins}. Substituting $U(\vec r,t)$ by
$A(t)U(\vec r)A^+(t)$ in (\ref{BF}), where $A$ is a SU(2) matrix, we obtain

\begin{equation}
\label{BFC}
L = - F_\pi \int d^3r \[ m - I\, Tr ( \partial_0 A 
\partial_0 A^{-1} ) \]^{3\over 2},
\end{equation}

\noindent  where $m$ and $I$, identified as being
the soliton mass and the inertia moment\cite{Adkins}, respectively, are
functionals written in terms of the chiral angle $F(r)$, which satisfies the topological
boundary conditions,$\,F(0)=\pi$ and $F(\infty)=0$. Here, we use the Hedgehog ansatz for $U$, i.e., $U=\exp(i\tau\cdot \hat{r}F(r))$.
The SU(2) matrix $A$ can be written as $A=a_0+i a_i \tau_i$, where $\tau_i$ are the Pauli matrices, leading
to the constraint

\begin{equation}
\label{primary}
\sum_{i=0}^{i=3} a_ia_i = 1.
\end{equation}

\noindent The Lagrangian (\ref{BFC}) can be written as a function
of the $a_i$ as

\begin{equation}
\label{BFCOL}
L = - {F_\pi\over 16}  \int d^3r \[  m - 2I \, \dot{a}_i\dot{a}_i  \]
^{3\over 2}.
\end{equation}

\noindent From the Eq. (\ref{BFCOL}) we can obtain the conjugate momenta,
given by

\begin{eqnarray}
\label{momen}
\pi_i = {\partial L \over \partial \dot{a}_i }
= {3 F_\pi\over 16} \,\,\dot{a}_i  
\int d^3r I \[ m - 2 I\,\dot{a}_k\dot{a}_k \]^{1 \over 2}.
\end{eqnarray}

\noindent The algebraic expression for the Hamiltonian is 
obtained applying the Legendre transformation, 
$ \, H = \pi_i\dot{a}_i - L \,$. 
However, in some situations, due to the momenta expression given in Eq.(\ref{momen}), 
it is not possible to write the conjugate Hamiltonian 
corresponding to
the Born-Infeld Skyrmion Lagrangian in terms of $\pi_i$ and $a_i$. 
An alternative procedure is to expand the original Lagrangian 
(\ref{BFCOL}) in collective coordinates. Thus, considering the 
binomial expansion variable \footnote{In the context of 
semi-classical expansion, 
it is expected that the product of $\dot{a_i}\dot{a_i}$ by the 
expression ${I\over m}$ given by the 
Euler-Lagrange equation, does not considerably modify this result.}
$\,\,{I\over m} \,\dot{a}_i\dot{a}_i \,\,$, the Lagrangian sum is 
given by

\begin{eqnarray}
\label{Lseries}
L = - M + A (\dot{a}_i\dot{a}_i) - B ({\dot{a}_i\dot{a}_i}^2)
+ \dots \,,
\end{eqnarray}

\noindent where $M={F_\pi\over 16} \int d^3r \,m^{3\over 2},\,\, A = {3F_\pi\over 16} \int d^3r \,\, I \sqrt{m},\,\, B={3F_\pi\over 32} \int d^3r \,\, {I^2 \over \sqrt{m}}$, and etc. In this step we would like to give physical argument that justifies this procedure. Even though not being a relativistic invariant model, we hope that the experimental results can be reproducible with a good accuracy when the soliton velocity is much smaller than the speed of light. From the relation given in ref.\cite{Weigel}, $A^+ {\partial_0 A}= i/2\sum_{k=1}^{k=3} \tau_k \omega_k$, where 
$\omega_k$ is the uniform soliton angular velocity, it is possible 
to show that $Tr[\partial_0 A \partial_0 A^+] = 2 \dot{a_i}\dot{a_i}
=\omega^2/2$. If we require that the soliton rotates with velocity 
smaller than {\it c}, then  $\omega r \ll 1$, leading to $\dot{a_i}
\dot{a_i}={\omega^2\over 4} \ll 1$, and consequently
$\dot{a_i}\dot{a_i} \ll 1$ for all space. Thus, these results explain 
our procedure.

In this manner, the Hamiltonian is obtained by using the Legendre transformation

\begin{eqnarray}
\label{Hamil1}
H & = & \pi_i \dot{a}_i - L \nonumber \\
& = & M + A (\dot{a}_i\dot{a}_i) -3B {(\dot{a}_i\dot{a}_i)}^2 + \dots\, .
\end{eqnarray}

\noindent Obtaining the canonical momenta  from Eq. (\ref{Lseries}), 
then writing the Lagrangian as $L=\pi_i\dot{a_i}-H$, and comparing with the 
expansion of the Lagrangian (\ref{Lseries}), it is possible to 
derive the expression of the Hamiltonian (\ref{Hamil1}) as

\begin{eqnarray}
\label{Hamilf}
H =  M + \alpha\, \pi_i\pi_i + \beta \, {(\pi_i\pi_i)}^2
 + \dots \,,
\end{eqnarray}

\noindent with $\, \alpha={1\over 4A}$ and $\beta={B\over 16A^4}$. We will truncate the 
expression (\ref{Hamilf}) in the second-order variable\footnote{
Due to the equation(\ref{momen}) together with the fact that 
$\dot{a_i}\dot{a_i} \ll 1\,\,$, we expect that terms like 
${(\pi_i\pi_i)}^3$ or higher order degree do not alter our
conclusion about the commutators of the quantum Born-Infeld 
Skyrmion.}, and we will use this approximate Hamiltonian to 
perform the quantization.

\section{Operatorial quantization at the\\ second-class level}

In this section the reduced Born-Infeld Skyrmion model will be
quantized using the Dirac method\cite{Dirac} and the symplectic
formalism\cite{FJ,JBW}. In order to apply the Dirac second-class
Hamiltonian method, we need to look for
secondary constraints, which can be calculated from the following 
Hamiltonian

\begin{eqnarray}
\label{formula1}
H_T = M+ \alpha \, \pi_i \pi_i + 
\beta \,(\pi_i\pi_i)^2 + \lambda_1 (a_ia_i-1),
\end{eqnarray}

\noindent where $\,\lambda_1\,$ is a Lagrangian multiplier. Using the iterative Dirac formalism 
we get the following second-class constraints

\begin{eqnarray}
\label{formula3}
\phi_1 &=& a_ia_i - 1 \approx 0, \\
\label{formula4}
\phi_2 &=& a_i\pi_i \approx 0.
\end{eqnarray}

After straightforward computations, the Dirac
brackets among the phase spaces variables are
obtained as

\begin{eqnarray}
\label{formula 7}
\{ a_i,a_j \}^*& = & 0,\nonumber\\
\label{formula 8}
\{ a_i,\pi_j \}^*& = & \delta_{ij} - a_i a_j ,\\
\label{formula 9}
\{ \pi_i,\pi_j \}^*& = &  a_j\pi_i - a_i\pi_j. \nonumber
\end{eqnarray}

\noindent Through the well known canonical quantization rule
$\{\,,\,\}^* \rightarrow -i\[\,,\,\]$, we get the commutators

\begin{eqnarray}
\label{formula10}
\[ a_i,a_j \]& = & 0,\nonumber\\
\[ a_i,\pi_j \]& = & -i \( \delta_{ij} - a_i a_j \) ,\\
\[ \pi_i,\pi_j \]& = & -i \( a_j\pi_i - a_i\pi_j \). \nonumber
\end{eqnarray}

\noindent These results show that the quantum commutators of the reduced Born-Infeld Skyrmion model are equal to the Skyrme
model\cite{Fujii,JAN} when the Lagrangian is expanded up to the
second-order term of the collective coordinates. This
completes the Dirac's quantization process. 

To implement the symplectic quantization procedure\cite{JBW}, let us consider the zeroth-iterative first-order Lagrangian

\begin{equation}
\label{formula11}
L^{(0)} = \pi_i\dot{a}_i - V^{(0)},
\end{equation}

\noindent where the potential $V^{(0)}$ is

\begin{equation}
\label{formula12}
V^{(0)} = M + \alpha\pi_i\pi_i + \beta(\pi_i\pi_i)^2 + \lambda (a_ia_i - 1),
\end{equation}

\noindent with the enlarged symplectic variables given by $\xi^{(0)}_\alpha=(a_j,\pi_j,\lambda)$. The symplectic tensor is

\begin{equation}
f^{(0)} = \left(
\begin{array}{ccc}
0           & -\delta_{ij} & 0 \\
\delta_{ij} &         0     & 0 \\
0           &         0     & 0
\end{array}
\right),
\end{equation}

\noindent where the elements of rows and columns follow the order: 
$a_i$, $\pi_i$, $\lambda$.
The matrix above is obviously singular, then it has a zero-mode that generates the following constraint 

\begin{equation}
\label{formula15}
\Omega^{(1)} = a_ia_i -1 \approx 0,
\end{equation}

\noindent where the potential $V^{(0)}$ is given by Eq.(\ref{formula12}).
Taking the time derivative of this constraint and 
introducing the result into the previous Lagrangian by means 
of a Lagrange multiplier $\rho$, we get a new Lagrangian $L^{(1)}$

\begin{equation}
\label{formula16}
L^{(1)} = (\pi_i + \rho a_i)\dot{a}_i - V^{(1)},
\end{equation}

\noindent where

\begin{equation}
\label{formula17}
V^{(1)} = M + \alpha\pi_i\pi_i + \beta (\pi_i\pi_i)^2.
\end{equation}
The matrix $f^{(1)}$ is then 

\begin{equation}
f^{(1)}=\left(
\begin{array}{ccc}
0           & -\delta_{ij} & -a_i \\
\delta_{ij} &         0     & 0 \\
a_i           &         0     & 0
\end{array}
\right),
\end{equation}

\noindent where rows and columns follow the order: $a_i$, $\pi_i$, $\rho$. 
The matrix $f^{(1)}$ is singular so it has a zero-mode, given by 

\begin{equation}
v^{(1)}=\left(
\begin{array}{ccc}
0 \\
a_i \\
-1
\end{array}
\right),
\end{equation}
that produces the constraint

\begin{equation}
\label{formula20}
\Omega^{(2)} = a_i\pi_i \approx 0.
\end{equation}

\noindent Here we must mention that these constraints,
Eqs.(\ref{formula15}) and (\ref{formula20}),
derived by the symplectic procedure are the same one obtained when the Dirac formalism is used. Following
the iterative symplectic process,
we get the new Lagrangian $L^{(2)}$, given by

\begin{equation}
\label{formula21}
L^{(2)} = (\pi_i + \rho a_i + \eta \pi_i)\dot{a}_i 
+ \eta a_i \dot{\pi}_i - V^{(2)},
\end{equation}

\noindent where $V^{(2)} = V^{(1)}$. 
The new enlarged symplectic variables are\break
$\xi^{(2)}_\alpha=(a_j,\pi_j, \rho,\eta)$, 
where $\rho$ and $\eta$ are Lagrange multipliers. 
The corresponding symplectic matrix $f^{(2)}$ is

\begin{equation}
f^{(2)}=\left(
\begin{array}{cccc}
0           & -\delta_{ij}  & -a_i   &  -\pi_i \\
\delta_{ij} &         0     &   0    &    -a_i \\
a_i         &         0     &   0    &     0   \\
\pi_i       &        a_i    &   0    &     0  
\end{array}
\right).
\end{equation}
The matrix $f^{(2)}$ is not singular, then it is identified as the symplectic tensor of the constrained theory. The inverse of $f^{(2)}$  gives the same Dirac brackets among the physical coordinates given in 
Eq.(\ref{formula 9}).

\section{Gauging the Born-Infeld Skyrmion model}

The Born-Infeld Skyrme model is a noninvariant model with  field dependent Dirac's brackets among the phase-space variables. Due to this, the quantization of the model is affected by operator ordering ambiguity. To overcome this problem at the commutator level, the model will be reformulated as a gauge invariant model. 
In this section, we will use the symplectic gauge-invariant  method proposed in section 4. To implement this scheme, the second-order Lagrangian that governs the dynamics of the Born-Infeld
Skyrmion model is reduced to its first-order form and an extra term $G(a_i,\pi_i,\theta)$ is introduced into the first-iterative Lagrangian, namely,

\be
\label{1000}
L^{(1)} = \pi_i\dot a_i + (a_i^2 - 1)\dot\eta - V^{(1)},
\ee
where $-\lambda\rightarrow\dot\eta$ and with $V^{(1)}$ as

\be
\label{1005}
V^{(1)} = V^{(0)}|_{(a^2-1=0)}= M + \frac{1}{4A} \pi_i^2 + \frac{B}{16A^4} \pi_i^4  - G(a_i,\pi_i,\theta),
\ee
The symplectic variables are $\xi_\alpha^{(1)}=(a_i,\pi_i,\eta,\theta)$ and the extra term, given by

\be
\label{1010}
G(a_i,\pi_i,\theta) = \sum_n^\infty{\cal G}^{(n)} (a_i,\pi_i,\theta),
\ee
satisfies the boundary condition,

\be
\label{1020}
G(a_i,\pi_i,\theta=0) = {\cal G}^{(0)} (a_i,\pi_i,\theta=0) = 0.
\ee
The corresponding symplectic matrix, computed as

\be
\label{1030}
f^{(1)} = \pmatrix{0 & -\delta_{ij} & 2a_i & 0\cr \delta_{ij} & 0 & 0 & 0\cr -2a_i & 0 & 0 & 0\cr 0 & 0 & 0 & 0},
\ee
is singular and has a zero-mode,

\be
\label{1040}
\nu^{(1)} = \pmatrix{0 & a_i & \frac 12 & 1}.
\ee
Following the prescription of the symplectic gauge-invariant formalism, giving in apendice A,
the gauge invariant Lagrangian obtained after
fourth iteration is

\be
\label{1130}
L^{(1)} = \pi_i\dot a_i + (a_i^2 - 1)\dot\eta  - V^{(1)}_{(4)},
\ee
\noindent where the symplectic potential $V^{(1)}_{(4)}$ is identified as being the invariant Hamiltonian 

\ba
\label{1140}
H &=& M + \frac{1}{4A} \pi_i^2 + \frac{B}{16A^4} \pi_i^4 - \left(\frac {1}{2A} + \frac{B}{4A^4} \pi_i^2\right) (a_i\pi_i)\theta\nonumber\\ &+& \left(a_i^2 + a_i^2\frac{B}{A^3}(a_i\pi_i)^2 + \frac{B}{2A^3}a^2\pi^2\right)\frac{\theta^2}{4A} - \frac{B}{4A^4}(a_i\pi_i)a^2\theta^3\nonumber\\ &+& \frac{B}{16A^4} a^4\theta^4.
\ea
To complete our gauge invariant reformulation, the infinitesimal gauge transformation is also computed. 
In agreement with the symplectic method, 
the zero-mode $\tilde\nu^{(1)}$ is the generator of the infinitesimal gauge transformation $(\delta{\cal O}=\varepsilon\tilde\nu^{(1)})$,

\ba
\label{1150}
\delta a_i &=& 0,\nonumber\\
\delta \pi_i &=& \varepsilon a_i,\nonumber\\
\delta\lambda &=& \dot\frac{\varepsilon}{2},\\
\delta\theta &=& \varepsilon,\nonumber
\ea
where $\varepsilon$ is an infinitesimal time-dependent parameter.

At this stage, some gauge fixing schemes will be implemented following the symplectic method that allows us to reveal a new and remarkable result. First, we require that $\chi_1=\theta=0$(unitary gauge) that reduces the gauge invariant model to the original model with the same Dirac's brackets among the phase-space variables $(a_i,\pi_i)$. The other one is

\be
\label{1170}
\chi_2=\lambda=0.
\ee

\noindent With this gauge we have another noninvariant description for the nonlinear model with canonical Dirac's brackets. In fact, the first-order Lagrangian becomes

\be
\label{1250}
L^{(1)} = \pi_i\dot a_i +\lambda\dot\rho - V^{(1)}_{(4)},
\ee
where the symplectic variables are $\xi_\alpha^{(1)} = (a_i,\pi_i,\lambda,\rho,\theta)$ and $V^{(1)}_{(4)}$ is

\ba
\label{1260}
V^{(1)}_{(4)} = V^{(0)}_{(4)}|_{\lambda=0} &=& M + \frac{1}{4A} \pi_i^2 + \frac{B}{16A^4} \pi_i^4 - \left(\frac {1}{2A}
+ \frac{B}{4A^4} \pi_i^2\right) (a_i\pi_i)\theta + \nonumber\\ &+& \left(a^2 + a^2 \frac{B}{A^3}(a_i\pi_i)^2 + \frac{B}{2A^3}a^2\pi^2\right)\frac{\theta^2}{4A} - \frac{B}{4A^4}(a_i\pi_i)a^2\theta^3 \nonumber\\
&+& \frac{B}{16A^4} a_i^4\theta^4.
\ea
The corresponding symplectic matrix, computed as

\be
\label{1270}
f^{(1)} = \pmatrix{ 0 & - \delta_{ij} & 0 & 0 & 0 \cr \delta_{ij} & 0 & 0 & 0 & 0\cr 0 & 0 & 0 & 1 & 0\cr 0 & 0 & -1 & 0 & 0 \cr 0 & 0 & 0 & 0 & 0},
\ee
is singular and has a zero-mode that produces a new constraint,

\ba
\label{1280}
\Omega_2 &=& \frac {1}{2A} (a_i\pi_i) + \frac{3B}{4A^4}(a_i\pi_i)a^2\theta^2 - \frac{B}{4A^4}a^4\theta^3 + \frac{B}{4A^4}\pi_i^2(a_i\pi_i) \nonumber\\ 
&-& \frac{1}{2A}a_i^2\theta - \frac{B}{2A^4}(a_i\pi_i)^2\theta - \frac{B}{4A^4}\pi^2 a^2\theta.
\ea
With the introduction of this constraint into the first-order Lagrangian $(L^{(2)})$, we have

\be
\label{1285}
L^{(2)} = \pi_i\dot a_i +\lambda\dot\rho  + \Omega_2\dot\beta - V^{(2)}_{(4)},
\ee
where $V^{(2)}_{(4)}= V^{(1)}_{(4)}|_{\Omega_2=0}$. After that, a nonsingular symplectic matrix is set up and the Dirac's brackets among the phase-space variables are identified as

\ba
\label{1290}
\lbrace a_i,a_j\rbrace^* &=& 0,\nonumber\\
\lbrace a_i,\pi_j\rbrace^* &=& \delta_{ij},\\
\lbrace \pi_i,\pi_j\rbrace^* &=& 0.\nonumber
\ea
>From $\Omega_2=0$, the $\theta$ variable can be determined as

\be
\label{1300}
\theta = \frac{1}{a^2}\left(a_i\pi_i + \frac{B}{2A^3}\pi_i^2(a_i\pi_i)\right).
\ee
Bringing back this result into the symplectic potential $V^{(2)}_{(4)}= V^{(1)}_{(4)}|_{\Omega_2=0}$ and collecting terms up to $\pi_i^4$, we obtain the following Hamiltonian,

\be
\label{1310}
H = V^{(2)} = M + \frac{1}{4A} \pi_iM_{ij}\pi_j + \frac{B}{16A^4} (\pi_iM_{ij}\pi_j)^2,
\ee
with the singular matrix $M_{ij}$ defined as

\be
\label{matriz1}
M_{ij}= \delta_{ij}-a_ia_j.
\ee
This Hamiltonian  will be used to perform the computation of the energy spectrum. At this stage it is important to notice that the noninvariant model has a hidden symmetry that could not be detected by the symplectic method, due to the inexistence of a gauge generator. In spite of this, the Hamiltonian (\ref{1310}) is invariant under the gauge infinitesimal transformations(\ref{1150}), because the matrix $M_{ij}$ has an eigenvector with eigenvalue null,

\be
\label{1320}
a_iM_{ij} = 0.
\ee
In the next section, the hidden symmetry will be investigated using
the gauge unfixing method.

\section{The gauge invariant Born-Infeld Skyrmion model}

In this section the hidden symmetry which underlies in the
Born-Infeld Skyrmion model
will be disclosed using the gauge unfixing Hamiltonian method\cite{MR},
reviewed in the appendix. This model has a set of second-class constraints, given in (\ref{formula3}) 
and (\ref{formula4}), that produces the nonvanishing Poisson bracket

\be
\label{B000010}
C = \lbrace\phi_1,\phi_2\rbrace = 2 a_ia_i = 2.
\ee

To obtain the first-class Hamiltonian in a systematic way we follow closely the procedure described in \cite{VT}. Initially, the
set of constraints are redefined as

\ba
\label{B000020}
\xi &=& C^{(-1)}\phi_1 = {1\over 2} a_ia_i - {1\over 2},\nonumber\\
\psi &=& \phi_2,
\ea
that generates the canonical Poisson bracket

\be
\label{B000030}
\lbrace\xi,\psi\rbrace = 1.
\ee
Subsequently, the Lagrange multipliers $u_a$ with $a=1,2$ which appears in the total Hamiltonian

\begin{eqnarray}
\label{total21}
H = H_c + u_1\xi + u_2\psi,
\end{eqnarray}
with

\begin{eqnarray}
\label{hcbi}
H_c= M + \alpha \pi_i\pi_i + \beta (\pi_i\pi_i)^2,
\end{eqnarray}
are determined as 

\ba
\label{B000040}
u_1 &=& -2\alpha\pi_i^2 - 4\beta (\pi_i^2)^2,\nonumber\\
u_2 &=& -2\alpha a_i\pi_i - 4\beta(a_i\pi_i)\pi_i^2,
\ea
just imposing that the constraints has no time evolution. 
To obtain the first-class system, we maintain only $\xi$ as gauge generator. At first, $\{\xi,H\} \neq 0$, i.e., $\xi$ and $H$, in
principle, do not satisfy a first-class algebra. Thus, the first-class Hamiltonian can be given by the following formula\cite{VT}

\begin{eqnarray}
\label{HF1}
\tilde{H} = H - \psi \{ \xi, H \} + {1\over 2!} \psi^2 
\{ \xi ,  \{ \xi, H \} \}
-  {1\over 3!} \psi^3 \{\xi, \{ \xi , \{ \xi, H \} \} + \dots,
\end{eqnarray}
with

\begin{eqnarray}
\label{ht2}
H= M+\alpha \pi_i\pi_i+\beta (\pi_i\pi_i)^2-
2\alpha (a_i\pi_i)^2 + 4\beta (a_i\pi_i)^2 (\pi_j\pi_j),
\end{eqnarray}
satisfying the first-class algebra

\be
\label{fc1}
\{\xi,\tilde{H}\} = 0.
\ee

At this point, we start to compute each term of the Hamiltonian (\ref{HF1}). The first one is

\begin{eqnarray}
\label{hb1}
\{ \xi, H \} = -2\alpha (a_i\pi_i) + 12\beta (a_i\pi_i) (\pi_i\pi_i)
+ 8\beta (a_i\pi_i)^3,
\end{eqnarray}
while the second and the remaining ones are given by

\ba
\label{hb2}
\{ \xi,\{\xi,H\}\}&=&-2\alpha+12\beta (\pi_i\pi_i) + 48\beta (a_i\pi_i)^2,\nonumber\\
\{\xi, \{ \xi,\{\xi,H\}\}\}&=& 120\beta (a_i\pi_i),\\
\{\xi,\{\xi,\{ \xi,\{\xi,H\}\}\}\}&=& 120\beta.\nonumber
\ea
The gauge invariant Hamiltonian is obtained as

\begin{eqnarray}
\label{hfirst}
\tilde{H}&=&M+\alpha (\pi_i\pi_i)+\beta(\pi_i\pi_i)^2-2\beta (a_i\pi_i)^2
(\pi_j\pi_j)-\alpha (a_i\pi_i)^2+\beta (a_i\pi_i)^4\nonumber\\
&=& M + \frac{1}{4A} \pi_iM_{ij}\pi_j + \frac{B}{16A^4} (\pi_iM_{ij}\pi_j)^2,
\end{eqnarray}
with the singular matrix $M_{ij}$ given in 
Eq.(\ref{matriz1}). 

It is easy to show that the gauge invariant Hamiltonian satisfies the noninvolutive algebra

\begin{eqnarray}
\label{hn}
\{ \xi, \tilde{H} \} = 0.
\end{eqnarray}
Due to this, the constraint $\xi$ is the gauge symmetry generator of the infinitesimal transformation

\ba
\label{S1}
\delta a_i &=& \lbrace a_i,\xi\rbrace = 0,\nonumber\\
\delta\pi_i &=& \lbrace \pi_i,\xi\rbrace = - \varepsilon a_i,
\ea
with $\varepsilon$ as an infinitesimal time-dependent parameter. Note that the Hamiltonian (\ref{hfirst}) is invariant under this infinitesimal gauge transformation because $a_i$ are eigenvectors of the phase-space metric $M_{ij}$ with null eigenvalues $(a_iM_{ij} = 0)$.

\section{The energy spectrum}

In this section, we will derive the energy levels.
Normally, these results were employed to obtain the baryons physical properties\cite{Adkins}. 
In this context, our perturbative approach plays an important role on the computation of the energy spectrum since the quartic term presenting in the Hamiltonian(\ref{Hamilf}) leads to an extra term.

In the second-class formalism the energy spectrum is obtained calculating the mean value of the quantum Hamiltonian, which reads as

\ba
\label{sp}
E = \langle\psi| \tilde{H} | \psi \rangle,
\ea
where $\tilde{H}= M + \alpha\, \pi_i\pi_i + \beta \, 
{(\pi_i\pi_i)}^2 $. $\tilde{H}$ is the quantum version of the second-class Hamiltonian, Eq.(\ref{Hamilf}). The eigenvectors of the quantum Hamiltonian $\tilde{H}$ are $| \psi \rangle={1\over N(l)}(a_1 + i a_2)^l=|polynomial\rangle \,$. These wave functions are also eigenvectors of the spin and
isospin operators, written in reference\cite{Adkins} as $ J_k={1\over 2}( a_0 \pi_k -a_k \pi_0 - \epsilon_{klm} a_i \pi_m )$  and $ I_k={1\over 2 } ( a_k \pi_0 -a_0 \pi_k- \epsilon_{klm} a_i\pi_m )$. The expression for $\pi_i$, satisfying the commutation relations, Eqs.(\ref{formula10}), is given by

\be
\label{m}
\pi_i= {1\over i} (\partial_i-a_ia_j \partial_j).
\ee
The algebraic expression for $\pi_i$ presents operator ordering problems. A possible choice, following the prescription of Weyl ordering\cite{Weyl}
(symmetrization procedure) is given by

\ba
\label{mw}
[p_i]_{sym}={1\over 6i} (6\partial_i
-a_i a_j\partial_i - a_i\partial_j a_j
- a_j a_i\partial_i - a_j\partial_j a_i\nonumber\\ 
-\partial_j a_i a_j
-\partial_j a_j a_i)\nonumber\\ \nonumber \\
=  {1\over i} \( \partial_i-a_i a_j \partial_j
- {5\over 2} a_i \).
\ea
Consequently, $\pi_j\pi_j$ symmetrized can be written as

\ba
\label{pi2}
{[\pi_j\pi_j]}_{sym}= -\partial_j\partial_j+{1\over2}\(
OpOp+2Op+{5\over4} \),
\ea
where $Op$ is defined as $ Op\equiv a_i\partial_i $.
The symmetrized second-class Hamiltonian operator is

\ba
\label{Hqsym}
[\tilde{H}]_{sym} = M+\alpha \[ -\partial_j\partial_j+{1\over2}\(
 OpOp+2Op+{5\over4} \)\]\nonumber\\
+\beta \[ [ -\partial_j\partial_j+{1\over2}\(
 OpOp+2Op+{5\over4} \)]\]^2.
\ea
Substitution of the expression (\ref{Hqsym}) in the mean value, (\ref{sp}), leads to the energy levels\footnote{
\noindent Note that the eigenvalues of the operator Op are defined by the following
equation:\\ $Op |polynomial\rangle = l |polynomial\rangle$.} 

\ba
\label{energy2}
E_l=_{phys}\langle\psi| \tilde{H} | \psi \rangle_{phys}=M+\alpha \[ l(l+2)+{5\over 4} \]
+\beta \[  l(l+2)+{5\over 4} \]^2.
\ea
Notice that these energy levels have a quartic extra term, indicating some
modifications on the calculation of the physical parameters, previously
obtained in the context of the SU(2) Skyrme model\cite{Adkins}. 
Furthermore, we remark that the adopted ordering scheme  produces 
a constant value 
on the energy levels formula. It is an important subject since different ordering schemes can lead to distinct physical results, as pointed out in Refs.\cite{Toda,JAN}.

In the first-class scenario the quantum Hamiltonian is

\ba
\label{qHf}
\tilde{H}=M+\alpha (\pi_i\pi_i)+\beta(\pi_i\pi_i)^2
-2\beta (a_i\pi_i)^2 (\pi_j\pi_j)-\alpha (a_i\pi_i)^2\nonumber\\+\beta (a_i\pi_i)^4.
\ea
$\tilde{H}$ is the quantum version of the first-class Hamiltonian, Eqs.(\ref{1310}). The quantization is performed, following the prescription of the Dirac method\cite{Dirac}, imposing that the physical wave functions are annihilated by the first-class operator constraint

\begin{eqnarray}
\label{qope}
\phi_1 | \psi \rangle_{phys} = 0,
\end{eqnarray}

\noindent where $\phi_1$ is

\begin{eqnarray}
\label{xi}
\phi_1= a_i a_i-1.
\end{eqnarray}

\noindent The physical states that satisfy (\ref{qope})
are

\begin{eqnarray}
\label{physical}
| \psi \rangle_{phys} = {1\over V } \, 
\delta(a_i a_i-1)\,|polynomial\rangle,
\end{eqnarray}
where {\it V } is the normalization factor. 
Thus, in order to obtain the spectrum of the theory, we take the scalar product, 
$_{phys}\langle\psi| \tilde{H} | \psi \rangle_{phys}\,$,
that is the mean value of the first-class Hamiltonian. 
We begin by calculating the scalar product

\begin{eqnarray}
\label{mes1}
_{phys}\langle\psi| \tilde{H} | \psi \rangle_{phys}=\nonumber \\
\langle polynomial |\,\,  {1\over V^2}  \int da_i\,\,
\delta(a^i a^i - 1)\,
\tilde{H}\,
\delta(a_i a_i - 1)\,\,
| polynomial \rangle,
\end{eqnarray}

\noindent where $\tilde{H}$ is defined in Eqs.(\ref{qHf}).  Note that due to $\delta(a_i a_i-1)$ in (\ref{mes1}) the scalar product can be simplified. Then, integrating over $a_i$, we obtain

\begin{eqnarray}
\label{mes2}
_{phys}\langle\psi| \tilde{H} | \psi \rangle_{phys}=\nonumber \\
\langle polynomial | M+\alpha (\pi_i\pi_i)+\beta(\pi_i\pi_i)^2
-2\beta (a_i\pi_i)^2 (\pi_j\pi_j)-\alpha (a_i\pi_i)^2\nonumber\\+\beta (a_i\pi_i)^4
| polynomial \rangle.
\end{eqnarray}

\noindent Here we would like to comment that the regularization delta function squared $\delta(a_i a_i - 1 )^2$ is performed using the delta relation, $(2\pi)^2\delta(0)=\\
\lim_{k\rightarrow 0}\int d^2x \,e^{ik\cdot x} =\int d^2x= V.$ 
In this manner,the parameter V is used as the normalization factor.
The Hamiltonian operator inside the kets, Eq.(\ref{mes2}), can be rewritten as

\begin{eqnarray}
\label{mes3}
_{phys}\langle\psi| \tilde{H} | \psi \rangle_{phys}=\nonumber \\
\langle polynomial | M + \alpha \[ p_k\cdot p_k \] 
+ \beta \[ p_k\cdot p_k \]^2| polynomial \rangle,
\end{eqnarray}
where $p_k=\pi_k - a_k (a_j\pi_j)$. 
The operators $\pi_k$ describe a free particle.
Then, the $p_k$ operators are identical to the canonical momenta obtained for the second-class theory. Consequently, the algebraic expression for the quantum Hamiltonian inside the scalar product(\ref{mes3}) is the same obtained in a second-class theory, Eq.(\ref{Hqsym}), naturally leading 
to the same energy levels, Eq.(\ref{energy2}). This important result shows the equivalence between the second-class collective coordinates Skyrme 
model and its first-class version. 

\section{Conclusions} 
The Born-Infeld Skyrmion Lagrangian has a nonconventional structure that allows to stabilize the soliton solutions without adding higher derivative 
order terms. However, due to the nonanalytical structure of Born-Infeld Skyrmion Lagrangian, a perturbative treatment becomes necessary. The expansion of the non-polynomial Born-Infeld Lagrangian in terms of dynamic variables is possible if we pay attention to the problem of breaking the relativistic invariance in the collective coordinates expansion. The contributions to the physical parameters due to the non-causal soliton solution have no physical relevance if the chiral angle F(r) satisfies the relation, $\lim_{r\rightarrow \infty}F(r) = 0 $, together with the fact that we impose that the soliton angular velocity be small, i.e., $ \omega \ll 1$. In views of this, the perturbative expression for the Hamiltonian could be truncated at the quartic-order term in the canonical momenta.

To obtain the quantum structure for the Born-Infeld Skyrmion model, the 
Dirac Hamiltonian method and the Lagrangian symplectic formalism were used. In these contexts, we verified that all constraints are second-class and the symplectic matrix is nonsingular, showing that is not expected to find symmetries. Afterward, the quantum commutators of the model were computed. These results are the same ones obtained for the conventional SU(2) Skyrme model. In spite of this, the energy levels with a quartic correction term together with an operator ordering scheme can change the baryons static properties, previously obtained for the usual SU(2) Skyrme model.

In order to disclose the hidden symmetry, two different approaches were used: the symplectic gauge-invariant scheme and the unfixing Hamiltonian formalism.
The usual directions\cite{FS,BT,IJMP} point out the enlargement of the phase space with WZ variables, and consequently the symmetries arise. In our work, those symmetries are revealed only on the original phase space, where the quantum structure was obtained using the Dirac first-class procedure. In this scenario, the energy levels were computed, reproducing the spectrum of the original second-class system. Our findings point out the consistency of those first-class conversion procedures and propose the gauge invariant version for the Born-Infeld Skyrmion model, dynamically equivalent to the usual SU(2) Skyrme model.

\section{ Acknowledgments}
We would like to thank A.G. Sim\~ao for critical
reading. This work is supported in part by FAPEMIG, Brazilian Research Council.  In particular, C. Neves would like to  acknowledge the FAPEMIG grant no. EX-00005/00.

\appendix
\section{The symplectic gauge-invariant method}

 There are several schemes to reformulate noninvariant models as gauge invariant theories in the literature. However, recently, some 
constraint conversion formalisms, based on the Dirac's method\cite{Dirac}, were developed using Faddeev's idea of phase-space extension with the 
introduction of auxiliary variables \cite{FS}. Among them, the Batalin-Fradkin-Fradkina-Tyutin(BFFT)\cite{BT} and the iterative\cite{IJMP} methods were powerful enough to be successfully applied to a great 
number of important physical models. Although these techniques share
the same conceptual basis \cite{FS} and treat constrained systems
following the Dirac process\cite{Dirac}, the implementation of the 
constraint conversion methods are different. Historically, both BFFT and the iterative methods
were applied in linear systems, such as chiral gauge
theories\cite{IJMP,many3}, in order to eliminate the gauge anomaly that hampers the quantization process.
In spite of the great success achieved by these methods, some ambiguities which appear in the constraint conversion process make these iterative methods a hard task to implement\cite{BN}. It happens because these formalisms are based on the 
Dirac's method. In this section, we reformulate noninvariant systems 
as gauge invariant theories
using a new technique which is not affected by those ambiguities
problems. This technique follows the Faddeev suggestion\cite{FS}
and is set up on a contemporary framework to handle noninvariant 
model, namely, the symplectic formalism\cite{FJ,JBW}. 

In order to systematize the symplectic gauge-invariant formalism, 
a general noninvariant mechanical model whose dynamics is governed by a 
Lagrangian ${\cal L}(a_i,\dot a_i,t)$(with $i=1,2,\dots,N$) is 
considered, where $a_i$ and $\dot a_i$ are the space and velocities variables respectively. Notice that this consideration does not lead to the loss of generality or of physical content. Following the symplectic method, the first-order Lagrangian, written in terms of the symplectic variables $\xi^{(0)}_\alpha(a_i,p_i)$ (with $\alpha=1,2,\dots,2N$), reads as
 
\begin{equation}
\label{2000}
{\cal L}^{(0)} = A^{(0)}_\alpha\dot\xi^{(0)}_\alpha - V^{(0)},
\end{equation}
where $A^{(0)}_\alpha$ is the one-form canonical momenta, $(0)$ indicates that it is the zeroth-iterative Lagrangian and, $V^{(0)}$, the
 symplectic potential. After that, the symplectic tensor, defined as

\begin{eqnarray}
\label{2010}
f^{(0)}_{\alpha\beta} = {\partial A^{(0)}_\beta\over \partial \xi^{(0)}_\alpha}
-{\partial A^{(0)}_\alpha\over \partial \xi^{(0)}_\beta},
\end{eqnarray}
is computed. Since this symplectic matrix is singular, it has a zero-mode $(\nu^{(0)})$ that generates a new constraint when contracted with the gradient of potential, namely,

\begin{equation}
\label{2020}
\Omega^{(0)} = \nu^{(0)}_\alpha\frac{\partial V^{(0)}}{\partial\xi^{(0)}_\alpha}.
\end{equation}
Through a Lagrange multiplier $\eta$, 
this constraint is introduced into the zeroth-iterative Lagrangian (\ref{2000}), generating the next one

\begin{eqnarray}
\label{2030}
{\cal L}^{(1)} &=& A^{(0)}_\alpha\dot\xi^{(0)}_\alpha - V^{(0)}+ \dot\eta\Omega^{(0)},\nonumber\\
&=& A^{(1)}_\alpha\dot\xi^{(1)}_\alpha - V^{(1)},
\end{eqnarray}
where

\begin{eqnarray}
\label{2040}
V^{(1)}&=&V^{(0)}|_{\Omega^{(0)}= 0},\nonumber\\
\xi^{(1)}_\alpha &=& (\xi^{(0)}_\alpha,\eta),\\
A^{(1)}_\alpha &=& A^{(0)}_\alpha + \eta\frac{\partial\Omega^{(0)}}{\partial\xi^{(0)}_\alpha}.\nonumber
\end{eqnarray}
The first-iterative symplectic tensor is computed and 
since this tensor is nonsingular, the iterative process stops and the Dirac's brackets among the phase-space variables are obtained from the inverse matrix. On the contrary, if the tensor is singular, a new constraint arises and the iterative process goes on.

After this brief review, the symplectic gauge formalism will be systematized. It starts after the first iteration with the introduction of
an extra term dependent on the original and Wess-Zumino(WZ) variable, $G(a_i,p_i,\theta)$, into the first-order Lagrangian. This extra term, expanded as

\begin{equation}
\label{2060}
G(a_i,p_i,\theta)=\sum_{n=0}^\infty{\cal G}^{(n)}(a_i,p_i,\theta),
\end{equation}
where ${\cal G}^{(n)}(a_i,p_i,\theta)$ is a term of order n in $\theta$, satisfies the following boundary condition, 
\begin{equation}
\label{2070}
G(a_i,p_i,\theta=0) = {\cal G}^{(n=0)}(a_i,p_i,\theta=0)=0.
\end{equation}
The symplectic variables are extended to contain also  the WZ variable $\tilde\xi^{(1)}_{\tilde\alpha} = (\xi^{(0)}_\alpha,\eta,\theta)$
 (with ${\tilde\alpha}=1,2,\dots,2N+2$) and the first-iterative symplectic potential becomes

\begin{equation}
\label{2075}
{\tilde V}_{(n)}^{(1)}(a_i,p_i,\theta) = V^{(1)}(a_i,p_i) - \sum_{n=0}^\infty{\cal G}^{(n)}(a_i,p_i,\theta).
\end{equation}
For $n=0$, we have

\begin{equation}
\label{2075a}
{\tilde V}_{(n=0)}^{(1)}(a_i,p_i,\theta) = V^{(1)}(a_i,p_i).
\end{equation}
Subsequently, we impose that the symplectic tensor 
($f^{(1)}$) be a singular matrix with the corresponding
zero-mode 

\begin{equation}
\label{2076}
\tilde\nu^{(1)}_{\tilde\alpha}=\pmatrix{\nu^{(1)}_\alpha & 1},
\end{equation}
as the generator of gauge symmetry. Due to this, all correction terms \break ${\cal G}^{(n)}(a_i,p_i,\theta)$ in order of $\theta$ can be
 explicitly computed. Contracting the zero-mode $(\tilde\nu^{(1)}_{\tilde\alpha})$ with the gradient of potential ${\tilde
  V}_{(n)}^{(1)}(a_i,p_i,\eta,\theta)$ and imposing that no more constraints are generated, the following differential equation is obtained

\begin{eqnarray}
\label{2080}
\tilde\nu^{(1)}_{\tilde\alpha}\frac{\partial {\tilde V}_{(n)}^{(1)}(a_i,p_i,\theta)}{\partial{\tilde\xi}^{(1)}_{\tilde\alpha}}&=&0,\nonumber\\
\nu^{(1)}_\alpha\frac{\partial V^{(1)}(a_i,p_i)}{\partial\xi^{(1)}_\alpha} - \sum_{n=0}^\infty\frac{\partial{\cal G}^{(n)}(a_i,p_i,\theta)}{\partial\theta}&=&0,
\end{eqnarray}
that allows us to compute all correction terms in order of $\theta$. For linear correction term, we have

\begin{equation}
\label{2090}
\nu^{(1)}_\alpha\frac{\partial V_{(n=0)}^{(1)}(a_i,p_i)}{\partial\xi^{(1)}_\alpha} - \frac{\partial{\cal
 G}^{(n=1)}(a_i,p_i,\theta)}{\partial\theta} = 0,
\end{equation}
while for quadratic one,

\begin{equation}
\label{2095}
{\nu}^{(1)}_{\alpha}\frac{\partial V_{(n=1)}^{(1)}(a_i,p_i,\theta)}{\partial{\xi}^{(1)}_{\alpha}} - \frac{\partial{\cal
 G}^{(n=2)}(a_i,p_i,\theta)}{\partial\theta} = 0.
\end{equation}
>From these equations, a recursive equation for $n\geq 1$ is proposed as

\begin{equation}
\label{2100}
{\nu}^{(1)}_{\alpha}\frac{\partial V_{(n-1)}^{(1)}(a_i,p_i,\theta)}{\partial{\xi}^{(1)}_{\alpha}} - \frac{\partial{\cal
 G}^{(n)}(a_i,p_i,\theta)}{\partial\theta} = 0,
\end{equation}
that allows us to compute each correction term in order of $\theta$. This iterative process is repeated successively until the
 equation (\ref{2080}) becomes identically null, consequently, the term $G(a_i,p_i,\theta)$ is obtained explicitly. At this stage, the
  gauge invariant Hamiltonian, identified as being the symplectic potential, is obtained as

\begin{equation}
\label{2110}
{\tilde{\cal  H}}(a_i,p_i,\theta) = V^{(1)}_{(n)}(a_i,p_i,\theta) = V^{(1)}(a_i,p_i) + G(a_i,p_i,\theta),
\end{equation}
and the zero-mode ${\tilde\nu}^{(1)}_{\tilde\alpha}$ is identified as being the generator of an infinitesimal gauge transformation

\begin{equation}
\label{2120}
\delta{\tilde\xi}_{\tilde\alpha} = \varepsilon{\tilde\nu}^{(1)}_{\tilde\alpha},
\end{equation}
where $\varepsilon$ is an infinitesimal time-dependent parameter. 

\section{The gauge unfixing Hamiltonian formalism}

The main idea of the unfixing gauge procedure is to consider half of
the total second-class constraints as gauge fixing
terms\cite{VT,HT} and the remaining as gauge generators of symmetry.
To obtain a first-class Hamiltonian in a systematic way we follow
closely the procedure described by Vytheeswaran in \cite{VT}.
To start we consider a system with two second-class constraints,
$\phi_1$ and  $\phi_2$, where the Poisson bracket is

\be
\label{A00030}
C = \{ \phi_1, \phi_2 \}.
\ee
Using this relation and redefining the second-class constraints as

\ba
\label{A00040}
\xi &\equiv& C^{-1} \phi_1,\nonumber\\
\psi &\equiv& \phi_2,
\ea
we have

\be
\label{A00050}
\{\xi, \psi\}= 1 + \lbrace C^{-1},\psi\rbrace C\xi,
\ee
so that $\xi$ and $\psi$ are canonically conjugate on the surface
defined by $\xi=0$.

The total Hamiltonian is

\begin{eqnarray}
\label{total2}
H = H_c + u_1\xi + u_2\psi.
\end{eqnarray}
To obtain the first-class system, we maintain only $\xi$ as a constraint
 relation. At first, $\{\xi,H\} \neq 0$, i.e., $\xi$ and $H$, in
principle, do not satisfy a first-class algebra. Thus, the first-class
Hamiltonian can be expressed by the formula given in \cite{VT}, reads as

\begin{eqnarray}
\label{HF}
\tilde{H} = H - \psi \{ \xi, H \} + {1\over 2!} \psi^2 \{ \xi ,
\{ \xi, H \} \}
-  {1\over 3!} \psi^3 \{\xi, \{ \xi , \{ \xi, H \} \} + \dots,
\end{eqnarray}
which satisfies the first-class condition

\begin{eqnarray}
\label{fc}
\{\xi,\tilde{H}\} =0.
\end{eqnarray}

\noindent The first-class Hamiltonian $\tilde{H}$ can be elegantly
 rewritten in a projection equation form, given by

\begin{eqnarray}
\tilde{H} = P H \equiv : \exp^{-\psi \xi} : H,
\end{eqnarray}

\noindent with $\psi$ respecting the ordering rule, that is, it should come before the Poisson bracket.
The procedure described above is an outline of a formalism that converts a
second-class system into first-class one without enlargement of the phase space.

\end{document}